\documentclass[twocolumn,showpacs,preprintnumbers,amsmath,amssymb]{revtex4}

\usepackage{graphicx}
\usepackage{mathptmx, textcomp}
\usepackage[latin1]{inputenc}
\begin{document}

\hyphenation{Fesh-bach}
\title{Observation of interspecies Feshbach resonances in an ultracold Rb-Cs mixture}

\author{K. Pilch$^1$, A. D. Lange$^1$, A. Prantner$^1$, G. Kerner$^2$, F. Ferlaino$^1$\footnote{Corresponding author; francesca.ferlaino@uibk.ac.at}, H.-C. N\"{a}gerl$^1$, R. Grimm$^{1,2}$}
\affiliation{ $^1$Institut f\"{u}r Experimentalphysik und
Zentrum f\"{u}r Quantenphysik, \\
Universit\"{a}t Innsbruck,  6020 Innsbruck, Austria\\
$^2$Institut f\"{u}r Quantenoptik und Quanteninformation,\\
\"{O}sterreichische Akademie der Wissenschaften, 6020 Innsbruck, Austria}

\date{\today}


\begin{abstract}
We report on the observation of interspecies Feshbach resonances in an
ultracold, optically trapped mixture of Rb and Cs atoms. In a magnetic field
range up to 300 G we find 23 interspecies Feshbach resonances in the lowest
spin channel and 2 resonances in a higher channel of the mixture.
The extraordinarily rich Feshbach spectrum suggests the importance of different
partial waves in both the open and closed channels of the scattering problem along
with higher-order coupling mechanisms. Our results
provide, on one hand, fundamental experimental input to characterize the Rb-Cs
scattering properties and, on the other hand, identify possible starting points
for the association of ultracold heteronuclear RbCs molecules.

\end{abstract}

\maketitle

\section{Introduction}
Over the last years ultracold atomic physics has entered the new terrain of
more complex and composite quantum systems, such as quantum-degenerate atomic
mixtures and molecular quantum gases.  Great efforts have been made  to produce
heteronuclear mixtures with several combinations of atomic species
\cite{Modugno2001bec, Hadzibabic2002tsm,
Roati2002fbq,Deh2008fri,Taglieber2008qdt} and isotopes
\cite{Schreck2001qbe,Honda2002odf,Papp2006ooh,Mcnamara2006dbf}. A rich variety
of phenomena can be addressed with such mixtures as a result of  different
interactions, masses, and optical and magnetic properties. Moreover,
two-species mixtures are very attractive systems for molecular physics
\cite{Ospelkaus2006uhm,Weber2008aou,Ni2008ahp}. A particular interest consists
in heteronuclear molecules in the rovibrational ground state as they carry a
large permanent electric dipole moment, allowing to produce strongly
interacting dipolar quantum gases \cite{baranov2008tpi}.

Rubidium and cesium, the two heaviest stable alkali species, are both well
established in Bose-Einstein condensate (BEC) experiments. While Rb represents the first atomic species
ever condensed \cite{Anderson1995oob}, cesium BEC had to await the development
of efficient optical trapping methods \cite{Weber2003bec, Kraemer2004opo}. The
individual two-body interaction properties of Rb and Cs are very well
understood as a result of extensive studies by Feshbach spectroscopy
\cite{Marte2002fri,Chin2004pfs,ChengRMP}. Cesium shows a quite unique scattering behavior because of a very
large background scattering length in combination with many Feshbach resonances
and pronounced high-order Feshbach coupling. A few experiments on  Rb-Cs
mixtures have been performed in magnetic traps
\cite{Anderli2005sca,Haas2007ssm,Harris2008mto}. The observation of rapid
thermalization between the two different species
\cite{Anderli2005sca,Haas2007ssm} points to a large interspecies background scattering
length \cite{Tiesinga2006dot}. Feshbach resonances in the Rb-Cs system have not
been observed so far. Knowledge on such resonances is desired both as
spectroscopic input to precisely determine the scattering properties of the
mixture and to enable experimental control of the interspecies interaction.

A special motivation for combining Rb and Cs is the production of a quantum gas
of polar molecules, as recently demonstrated for a Bose-Fermi mixture of bosonic Rb and
fermionic K atoms \cite{Ni2008ahp}. As the Rb-Cs mixture is a Bose-Bose system,
the molecules would be bosons and could eventually form a condensate of
polar molecules \cite{baranov2008tpi}. For studying dipolar phenomena in such a system, RbCs
molecules in  the rovibrational ground state are expected to carry a
comparatively large electric dipole moment of 1.25 debye \cite{Kotochigova2005air}. For this application,
Feshbach resonances have proven an efficient tool to associate molecules and
thus to serve as a gateway from the atomic into the molecular world
\cite{ChengRMP,ferlaino2008cmt}. The Feshbach molecules are then transferred by
optical Raman processes into the rovibrational ground state \cite{Danzl2008qgo,
Lang2008utm,Ni2008ahp}.

In this Article, we report on Feshbach spectroscopy in an optically trapped
mixture of $^{87}$Rb and $^{133}$Cs atoms. After several optical cooling
stages, our ultracold mixture is loaded into an optical dipole trap (Sec.\,II).
We perform the Feshbach spectroscopy in two different spin state combinations
(Sec.\,III), exploring a magnetic field range of up to 300G. We find  a total
of 23 interspecies resonances for Rb and Cs atoms in their
lowest spin states. The extraordinarily rich spectrum indicates that dipolar
interactions, such as the second-order spin-orbit interaction, play an
important role for the scattering properties of the mixture (Sec.\,IV). Finally,
we discuss prospects to associate heteronuclear Feshbach molecules (Sec.\,V).

\section{Preparation of an optically trapped R\lowercase{b}-C\lowercase{s} mixture}

The preparation of an ultracold optically trapped Rb-Cs mixture employs several
stages of cooling and trapping acting simultaneously on both species. The
Rb and Cs atoms are first loaded into a two-color
magneto-optical trap (MOT) from a Zeeman slowed atomic beam. The mixture is
further cooled by applying two-color degenerate Raman sideband cooling (DRSC),
which also polarizes the atoms into their lowest spin state. The sample is then
loaded into a levitated optical dipole trap, which is used to probe
heteronuclear Feshbach resonances.

\subsection{Laser cooling}

The Rb and Cs atoms are initially emitted from a double-species oven operating
at a temperature of 100$^\circ$C and 80$^\circ$C for Rb and Cs, respectively
\cite{KarlThesis}. The average longitudinal velocity of Rb (Cs) is about 260
(210)~m/s. The two-species atomic beam is slowed down by using the Zeeman
slowing technique. The Zeeman-slower field is optimized  for Cs, but it works
well also for Rb because of the similar properties of the two species. The
decelerated atoms are then collected into a two-color MOT, simultaneously
confining and cooling the two species. The MOTs operate in a standard way, i.e.
for Rb on the $5 S_{1/2}(F\!=\!2)\!\to\!5 P_{3/2}(F'\!=\!3)$ atomic transition
and for Cs on the $6 S_{1/2}(F\!=\!4) \to 6 P_{3/2}(F'\!=\!5)$ transition. The
MOT beams for Rb and Cs are first superimposed on dichroic mirrors and then
sent through a single-mode optical fiber to ensure optimum spatial overlap of
both field components.  We use broadband polarization optics suitable for both
the Rb and the Cs wavelengths. After 10~s of MOT loading, we switch off the
Zeeman-slower light and we apply a compression stage for 15ms. This is
performed by ramping up the magnetic gradient field used for the MOT by a
factor of two, up to a maximum gradient of 14~G/cm along the symmetry axis, and
we simultaneously  detune the Rb (Cs) MOT light from about -10 (-8) MHz to -110
(-80) MHz. At the end of the compression phase, we switch off the MOT beams and
the magnetic gradient field.

The loading efficiency of a two-species MOT can be strongly affected by
inelastic interspecies collisions. For the Rb-Cs combination strong loss caused
by interspecies light-assisted collisions has been reported in
Ref.\,\cite{Harris2008mto}. Such losses crucially depend on the spatial overlap
between the species and  can be suppressed by introducing a relative
displacement. However a good spatial overlap is required to finally load the
mixture into the optical dipole trap.

To meet both requirements, we make use of a spatial displacement of the trapped
cloud induced by the radiation pressure of the Zeeman-slowing field and
controlled by fine-tuning its frequency. Subsequently, the  two atomic clouds
overlap during the compression stage, in which the Zeeman field is off.

For optimized loading of both MOTs, we measure about $5\times 10^8$ atoms for
each species at a temperature of about $ 40 \ \mu$K. For some experiments
(Sec.\,II.B) it is desirable to introduce an imbalance between the atom number
of the two species. This can be done by selectively reducing the power of the
respective MOT-light component.

\begin{figure}
\includegraphics[width=2.8in]{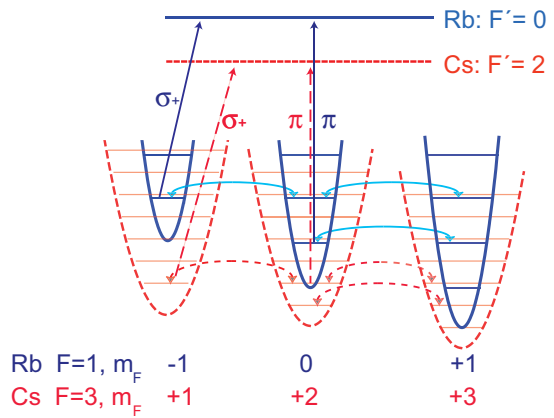}
\caption{color online. Illustration of the two-color DRSC technique applied
simultaneously to the Rb (blue solid lines) and Cs (red dashed lines) atoms.
For each species, a cycling transition consists of a two-photon Raman
transition  (indicated by the curved arrows) between vibrational levels of $m_F$-dependent
harmonic micro-traps followed by optical pumping to the lowest
Zeeman sublevel (not shown). Rb and Cs atoms accumulate in the vibrational
ground state of the $m_F=1$ and $m_F=3$ levels, respectively, which, as dark
states, are essentially decoupled from the light field.} \label{Fig1}
\end{figure}

We further cool the compressed mixture by applying a two-color version of DRSC.
This technique has the advantage to provide lower temperatures and higher
phase-space densities as compared to simple molasses cooling. Furthermore DRSC
polarizes the two species in their respective lowest spin states, labeled as
Rb$| 1, 1\rangle$ for Rb and Cs$| 3, 3\rangle$ for Cs, where the first number
indicates the total spin quantum number $F$ and the second its projection $m_F$
along the magnetic field axis. In single-species experiments, DRSC has been
proven a powerful method to cool and spin polarize individually both Cs
\cite{Kerman2000bom, Treutlein2001hba} and Rb atoms \cite{Weid2005}.  Here we
extend this technique to the case of an atomic mixture. The basic concept of
this cooling scheme adapted for double-species operation is illustrated in
Fig.\,\ref{Fig1}. The blue solid lines refer to the Rb and the red dashed lines
to the Cs case. The Rb and Cs atoms are simultaneously captured in two
independent three-dimensional optical lattices. The lattice light for the two
species is overlapped using dichroic mirrors. We produce the lattice using a
four-beam configuration \cite{Grynberg1993, Kraemer2004opo}, with total power
of about 28~mW and 20~mW for Rb and Cs, respectively. The waist of each beam is
about 1~mm. For Cs atoms in the lowest hyperfine manifold (F=3), the lattice
light is red-detuned  by 9.2 GHz from the $6 S_{1/2}(F\!=\!3) \to 6
P_{3/2}(F'\!=\!2)$ transition \cite{ED1}. The Rb lattice is red-detuned by
about 18~GHz with respect to the $5 S_{1/2}(F\!=\!1) \to 5 P_{3/2}(F'\!=\!0)$
transition. We observe that the lattice light for one species has negligible
effect on the other species. The lattices provide two independent arrays of
harmonic microtraps.

We apply a small homogenous magnetic field to Zeeman shift the harmonic
microtraps in such a way that the $(m_F,\nu)$ state is nearly degenerate with
the $(m_F+1,\nu+1)$ state, where $\nu$ refers to vibrational quantum number.
Once this condition is reached, the lattice light couples states with
$(m_F,\nu)$ and $(m_F+1,\nu+1)$ via a two-photon Raman transition. Typical
magnetic field values are about 500~mG. To fulfill simultaneously the
degeneracy condition for both Rb and Cs atoms, we first optimize on the sample of Cs atoms
and we then appropriately tune the power of the Rb lattice.

Two polarizing laser beams are applied to optical pump the two species into
their lowest spin states, Rb$| 1, 1\rangle$ and Cs$| 3, 3\rangle$. The optical
pumping fields are resonant with the $F=1\!\to\!F'=0$ and the $F=3\!\to\!F'=2$
transitions for Rb and Cs, respectively.  A mixed $\sigma^+/\pi $ polarization
of this light is necessary to assure that the atoms are driven towards the dark
states $(m_F\!=\!3,\nu\!=\!0)$ for Cs and $(m_F\!=\!1,\nu\!=\!0)$ for Rb. After
10~ms of DRSC, we  ramp down adiabatically the lattice power. For maximum
loading of both species, we typically measure $10^8$ atoms for each species at
temperatures of about $3 \ \mu$K and $2 \ \mu$K for Rb and Cs, respectively.

\begin{figure*}
\includegraphics[width=6.5in]{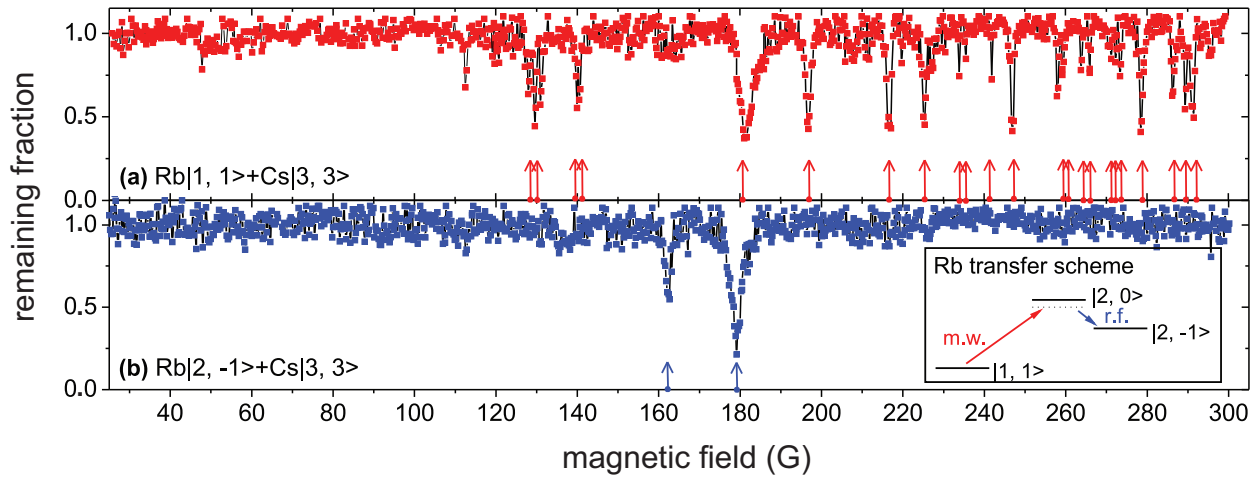}
\caption{color online. Observation of heteronuclear Rb-Cs Feshbach resonances
by loss spectroscopy. The  spectrum is recorded for Cs atoms,
being the minority component of the mixture. The remaining fraction represents
the number of Cs atoms after a 5-s hold time
in the dipole trap normalized to the corresponding average number detected off
any resonance.
The arrows indicate the positions of the interspecies Feshbach resonances. Other loss features correspond to
homonuclear Cs Feshbach resonances. (a) shows the loss spectrum for the Rb$| 1,
1\rangle$ + Cs$| 3, 3\rangle$ channel. (b) displays the Rb$| 2, -1\rangle$ +
Cs$| 3, 3\rangle$ channel. The inset  illustrates the transfer scheme applied
to populate the Rb$| 2, -1\rangle$ state.}\label{FR}
\end{figure*}

\subsection{Preparation in optical dipole trap}
\label{OT}

After the two-color DRSC stage, we load the atoms into a large-volume levitated
dipole trap \cite{Han2001lac}. Our choice of this trap configuration is
motivated by two main reasons. First, the trap acts as a perfect spin filter as
the levitation condition is only fulfilled for a specific magnetic substate.
Therefore our mixture is free of spin impurities and corresponding ambiguities
in spin channels of observed Feshbach resonances. Second, such a trap setup has
the advantage to cancel the influence of the gravitational force using
magnetic levitation. It is routinely employed in several experiments to produce
a BEC of Cs atoms \cite{Kraemer2004opo, Weber2003bec, gustavsson2008coi,
hung2008aec} and therefore holds prospects for future experiments on
two-species BECs of Rb and Cs.

In our experiment, the optical field is generated by an  ytterbium fiber laser
(IPG YLR-100) with a central wavelength of about 1070~nm and a linewidth of
about 3~nm. The dipole trap is produced by two horizontally propagating laser
beams crossing in the horizontal plane at a nearly right angle \cite{ED2}. Each
beam has a power of about 23~W and a waist of about $500\ \mu$m. The Rb atoms
experience a confining potential that is a factor of 1.7 shallower than the one
for Cs. This is due to the larger detuning and the lower optical polarizability
of Rb. For our typical parameters, the trap depth for Rb (Cs) atoms is $12 (20)
\ \mu$K.

A magnetic field gradient of $31.2$ G/cm is applied along the vertical
direction to hold the atoms against gravity, similarly to
Ref.\,\cite{Weber2003bec}. The levitation condition is simultaneously fulfilled
for both Rb and Cs atoms in their absolute ground states as a result of a
fortuitous coincidence in the  magnetic-moment-to-mass ratios of Rb and Cs.

The mixture is held in the optical trap for 2~s to allow for thermalization and
plain evaporation. In single-species experiments, Rb atoms have a lifetime of
about 10~s in the dipole trap. A shorter lifetime of about 3~s is measured for
Cs atoms \cite{ED3}.

We observe that Cs and Rb atoms rapidly thermalize during the evaporation via
sympathetic cooling, which points to a large cross section for interspecies
elastic collisions. Because of the different trap depths, the evaporative
cooling preferentially leads to losses of Rb atoms.  To produce a mixture  with
equal atom number, we start with a Cs MOT that has only $10^7$ atoms. In the
dipole trap we then finally measure $6 \times 10^5$ atoms for each species at a
temperature of  about $3 \ \mu$K and  peak density of a few $10^{10}$cm$^{-3}$.
By further reducing the MOT loading, we can also produce an imbalanced mixture
with much more Rb than Cs atoms. Note that for the present application of
Feshbach spectroscopy it was not necessary to optimize the mode matching
between the Raman cooled clouds and the dipole trap, resulting in a low loading
efficiency of the trap \cite{ED4}.

For Feshbach spectroscopy experiments, it is favorable to increase the atomic
density of both species since the losses occur as a result of interspecies
collisions. We compress the mixture by superimposing an additional  tightly
focused laser beam, which acts as a ``dimple trap'' \cite{Weber2003bec,
Kraemer2004opo}. The optical field is generated by a second fiber laser with central
wavelength of about 1064~nm and linewidth of about 1~nm. The beam has a power
of about 300~mW and a waist of $38 \ \mu$m. The dimple trap depths are $18 \
\mu$K and $31 \ \mu$K for Rb and Cs, respectively. With the use of this
additional optical field, we typically increase the peak density of both
species by an order of magnitude to about $5\times 10^{11}$ cm$^{-3}$
at a final temperature of about $7 \ \mu$K.

\section{ R\lowercase{b}-C\lowercase{s} Feshbach resonances }

A Feshbach resonance manifests itself in an enhancement of trap loss \cite{
ChengRMP}. In the energetically lowest spin channel, such losses can be fully
attributed to three-body collisions.  In our mixture, three-body collisions
involving one Cs and two Rb atoms and collisions with one Rb and two Cs atoms are both
possible. The rate coefficients for these two processes can in general be
different, with one process dominating over the other. The visibility of
heteronuclear resonances can be enhanced by creating an imbalance between the
atom numbers of the two species \cite{Wille2007eau, Deh2008fri}. In this way,
one species can be  used as a probe and the other one as a collisional bath.
For large enough atom imbalance, the minority component will be fully
depleted, while the majority component will remain nearly unaffected. By a
selective control of the MOT loading, we create a majority species (either Rb
or Cs) and a minority species (thus either Cs or Rb) with an imbalance ratio of
about $1:10$.

\subsection{Energetically lowest spin channel}

We perform Feshbach spectroscopy in the absolute ground state of the two
species, i.\,e.\,in their lowest spin-state combination Rb$| 1, 1\rangle$ + Cs$|
3, 3\rangle$. We search for heteronuclear resonances by performing a magnetic
field scan between 20 and 300 G with a step size of 250~mG.  At low magnetic
fields our scan is limited by the horizontal forces induced by the magnetic
levitation field, which pull the atoms out of the trap \cite{Weber2003bec}. A
scan consists of more than 1000 experimental cycles of about 30~s duration
each. In each cycle, we let the mixture evolve at a specific value of the
homogeneous magnetic field for 5~s. In the next cycle, we change the magnetic
field value in random order and we repeat the measurement. To determine the
atom number of the minority species, we switch off the optical dipole trap,
recapture the remaining atoms into the MOT, and record their fluorescence
signal. We trace the full loss spectrum using Cs atoms as probe; for some resonances
we also perform consistency checks by monitoring the loss of Rb atoms prepared
as a minority species.

Figure \ref{FR}(a) shows the complete  loss spectrum choosing Cs as the probe.
Each point corresponds to an average of 5 to 10 different measurements.  At
specific values of the magnetic field, the Cs atom number decreases by at least
$20\%$. Since the mixture contains an order of magnitude less Cs than Rb atoms,
the latter species exhibits very small relative losses. To avoid any possible
confusion with homonuclear resonances, we have repeated the magnetic field scan
using samples of either pure Cs or pure Rb atoms. The loss features, indicated
by the arrows in Fig.\,\ref{FR}(a), only appear using a two-species mixture,
while the others can be attributed to Cs Feshbach resonances
\cite{Chin2004pfs}.

In this way, we identify  23 interspecies Rb-Cs Feshbach resonances in the
absolute ground state combination of the two species.

Two main mechanisms can contribute to a broadening of the loss resonances.
First, the levitation field gives rise to an inhomogeneous broadening, which we
estimate to be of about 70~mG. Second, the finite temperature of the mixture
leads to a broadening on the order of 100~mG \cite{ED5}. As a result of both
broadening contributions, our resolution limit is not better than 100~mG.
Therefore we extract widths from our experimental data only for those
resonances that appear wider than this resolution limit. To determine the full
widths at half maximum, $\Delta B$, we use Lorentzian fits. The measured
positions, widths, and loss fractions are summarized in Table~\ref{resonances}.

\begin{table}[ht]
\caption{Feshbach resonances in collisions between $^{87}$Rb and $^{133}$Cs in
a range from 20 to 300G. The locations, the widths $\Delta B$  and the fractional losses of the Feshbach
resonances in both channels Rb$| 1, 1\rangle$ + Cs$| 3, 3\rangle$ and Rb$| 2,
-1\rangle$ + Cs$| 3, 3\rangle$ are listed.} \label{resonances}\centering
\begin{ruledtabular}
\begin{tabular}{c | c c c c c c}
& B (G)   & loss (\%) & $\Delta B$ (G)& B (G)& loss (\%) & $\Delta B$ (G)  \\
\hline
Rb$|1, 1\rangle$                                       &128.0  & 40 &  -&258.0  & 55   &   0.4  \\
+ Cs$|3, 3\rangle$                                     &129.6  & 60 & - &259.3  & 30    & 0.5\\
                                                      &140.0  & 50 &  -&263.8  & 30   & 0.5\\
                                                      &140.5  & 50 &  -&265.9  & 30   & -\\
                                                      &181.6  & 70 & 3.1 &271.2  & 25   &-\\
                                                      &196.8  & 60 & 1.2 &272.3  & 25   &0.3\\
                                                      &216.7  & 60 & 1 &273.4  & 30   &0.4\\
                                                      &225.3  & 60 & 1 &278.4   & 65  &0.95\\
                                                      &233.9  & 30&  - &286.2  & 45   & 0.8\\
                                                      &235.5  & 25&  - &289.4  & 50   &0.5\\
                                                      &241.9  & 35& -  &291.5  & 55   &1.1\\
                                                      &246.9  & 65& 0.7  &        &\\
&&&&&\\
Rb$|2,-1\rangle$ & 162.3 & 50 & 1.4 & &  & \\
+ Cs$|3,3\rangle$&179.1  & 80 & 2.8 & &  & \\
\end{tabular}
\end{ruledtabular}
\end{table}

\subsection{Higher spin channel}

In a second set of experiments, we perform  Feshbach spectroscopy in a higher
spin channel. We keep the Cs atoms in the absolute ground state because of
strong two-body dipolar losses occurring for Cs-Cs collisions in any higher
spin state \cite{Odelin1998ibe}. The Rb$| 2, -1\rangle$ state  is the only
other state simultaneously fulfilling the levitation condition. Atoms in the
Rb$|2, -1\rangle$ + Cs$| 3, 3\rangle$ channel are stable against spin-exchange
collisions that do not involve a change in $F$ as the relaxation channels are
energetically closed. Two-body hyperfine-changing collisions are energetically
possible and allowed under the conservation of the total orientation quantum
number, but we do not observe any significant background loss that exceeds the
observations in the other spin channel. Loss at Feshbach resonances may however
include resonantly enhanced two-body loss.

We start  with Rb atoms prepared in the Rb$| 1, 1\rangle$ state  as described
before. Then we apply a two-photon transfer scheme, which combines microwave
and radio-frequency excitation. The inset of Fig.\,\ref{FR}(b) schematically
shows the working principle of the transfer scheme. The microwave frequency is
detuned by few MHz with respect to the Rb$| 1, 1\rangle \rightarrow$ Rb$| 2,
0\rangle$ transition. The radio-frequency signal couples the Rb$| 2, 0\rangle$
and the Rb$| 2, -1\rangle$ state with the corresponding detuning that leads to
the two-photon resonance \cite{ED6}.

The pulse sequence has a duration of 400~ms during which the magnetic field
value is set at 20~G. Using this scheme, we incoherently transfer almost 50$\%$
of the ground state Rb atoms into the excited state \cite{ED7}. The remaining
atoms in the Rb$| 1, 1\rangle$ state  are selectively removed from the trap by
applying a short light pulse resonant with the $F=1\!\to\!F'=2$ transition,
pumping the atoms into different magnetic sublevels of the $F=2$ upper
hyperfine state. Those atoms which do not end up in the target state Rb$| 2,
-1\rangle$ leave the trap as the levitation condition is not fulfilled. In this
way spin impurities can be completely avoided.

To probe Feshbach resonances, we again perform a magnetic field scan, as
previously described. As shown in Fig.\,\ref{FR}(b), we observe two resonant
loss features. The positions, the widths, and the loss fractions are listed in
Table~\ref{resonances}. The observation of just two resonances in this higher
spin channel stands in contrast to the multitude of resonances observed in the
energetically lowest  channel.

\section{Discussion}
The interspecies scattering behavior of the ultracold Rb-Cs system is still an open issue.
The only piece of information available prior to the present work stems from
experiments in magnetically trapped Rb-Cs mixtures
\cite{Anderli2005sca,Haas2007ssm}.  The fast thermalization observed between Rb and Cs,
both being in their doubly-polarized states,  points to a large elastic cross section.
A theoretical analysis \cite{Tiesinga2006dot} has identified two possible values
of the interspecies triplet scattering length,
$700_{-300}^{+700} a_0$ and $176 _{-2}^{+2} a_0$,  the first value being the
more likely one \cite{Tiesinga2006dot}. Information on the scattering behavior
associated with the singlet potential and on the near-threshold molecular structure
is totally absent. Some knowledge
on deeply bound states  is available from photoassociation experiments, in which a
sample of RbCs molecules was produced in the rovibrational ground state
\cite{Sage2005opu}.

Our Feshbach spectroscopy opens up a way to substantially improve our
understanding of the interspecies scattering properties and the near-threshold
molecular structure of the RbCs system. The resonance positions provide
valuable experimental input to determine the relevant parameters of quantum
collisional models, such as the asymptotic bound state model
\cite{Stan2004ofr,Wille2007eau}  and coupled channels calculations
\cite{Stoof1988sea}.  Such scattering models have been  successfully applied to
the case of lighter combinations,  e.\,g.\,the Li-K \cite{Wille2007eau} and the
Rb-K \cite{Ferlaino2006fso,klempt2007krb} mixture. However, the resonance
assignment for heavier combinations, such as the Rb-Cs system, is much more
challenging  because of a larger hyperfine energy, a smaller vibrational
splitting in the molecular potentials, the spin-spin interactions (second order
spin-orbit and magnetic dipole interaction), and odd partial-wave contributions
to the scattering of non-identical bosons. These effects  result in an increase
of the density of the relevant bound states and in a mixing of different
molecular levels.

The Feshbach spectrum observed for the Rb-Cs system reflects
a variety of near-threshold molecular states and appears to be much richer than the
spectra observed for lighter mixtures \cite{Stan2004ofr,Inouye2004ooh,Ferlaino2006fso,Deh2008fri,Wille2007eau}.
In the Rb-Cs combination, the total hyperfine energy  of about $h\times$16~GHz substantially
exceeds the near-threshold vibrational splitting,
resulting in a mixing of at least five vibrational levels \cite{servaaspc}.
A large second-order spin-orbit interaction, which  mixes states with different angular
momenta $\ell$ up to $\Delta \ell$=4, is likely.
In particular,  Rb and Cs atoms colliding in an $s$ wave may couple to $s$-, $d$-, and $g$-wave
RbCs molecular states via Feshbach resonances. The important role of higher-order
coupling effects is well known from ultracold interactions in cesium samples \cite{Chin2000hrf}.
It has, for example, been demonstrated by populating molecular states
with high orbital angular momentum \cite{Mark2007sou,Mark2007siw,Knoop2008mfm}.
In a mixture,  collisions  between non-identical
particles can also take place in odd partial waves.
In our temperature regime ($T$=7$\mu$K), $p$-wave collisions may be particularly important
as the height  of the corresponding centrifugal potential barrier is only about $k_{\rm B}\times 55\mu$K.
The $p$-wave resonances can be induced by $p$-wave molecular states, or also through
higher-order couplings  by $f$-wave or $h$-wave states.

\begin{figure}
\includegraphics[width=2.8in]{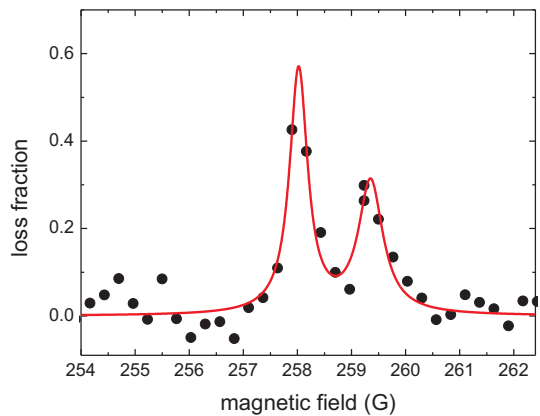}
\caption{color online. Example of a Rb-Cs resonance pair observed in the Rb$|1,
1\rangle$  + Cs$|3, 3\rangle$ channel. The circles represents a subset of the data shown
in Fig.\,\ref{FR} and the solid line is a fit to the data using a double Lorentzian profile. The resonance splitting is about 1.3 G.}
\label{Fig3}
\end{figure}

A $p$-wave resonance can lead to a doublet structure   because of the
spin-spin dipole interaction, which splits the resonance into two components based on
the partial-wave projection onto the magnetic field axis, $m_\ell =0$ and
$|m_\ell|=1$ \cite{Ticknor2004mso}. In our Feshbach scan, we observe
several resonances that come in pairs. Such a feature could be  fortuitous but could also be
the evidence of the $p$-wave character. In particular, four resonance pairs are
candidates for $p$-wave resonances, namely the resonances near 140, 234, 259, and 273~G.
We typically observe a splitting of about 1~G, which
would  indicate a strong spin-spin interaction between Rb and Cs atoms.
Doublets with a splitting of this order have been predicted for the specific
Rb-Cs system \cite{svetlanapc}.  Figure \ref{Fig3} gives an example of such a doublet
structure. Here, the splitting between the two peaks is about 1.3~G.

First calculations, based both on the asymptotic bound state model
and coupled channels theory, confirm the extraordinary richness of the Rb-Cs spectrum \cite{servaaspc,svetlanapc}.
Coupled channel calculations also show the importance of second-order spin-orbit
and spin-spin interaction and predict doublet structures of $p$-wave states \cite{svetlanapc}.
However, further refinements of the models are needed to reach an assignment of the
observed resonances.

\section{Outlook}
The observation of a variety of Feshbach resonances opens up several intriguing
avenues of research
both with tunable mixtures and heteronuclear molecules.
The production  of a tunable double BEC will allow to access complex quantum
phenomena, including phase separation, selective collapse, and vortex formation
\cite{Roberts2001cco, Tiemmermans1998pso,Garcia2000sau,Zaccanti2006cot}.
With the additional use of an optical lattice, the superfluid-to-Mott insulator
transition \cite{Lewenstein2004abf,Thalhammer2008dsb,bloch2008mbp} and repulsively bound pairs with an imbalance of masses
\cite{Winkler2006rba} can be studied. Heteronuclear resonances are also relevant
in few-body physics because of the possible enhancement of the observability of
the Efimov effect \cite{Kraemer2006efe, dincao2006eto}.

On the molecular side, the observed Feshbach resonances can be used as an entrance gate
for association of molecules of different character.
We expect that  RbCs molecules can be, for
instance, associated by ramping the magnetic field value across any of the
narrow and broad interspecies Feshbach resonances \cite{Herbig2003poa,Regal2003cum}.
Two particularly interesting directions can be pursued with heteronuclear molecules:
universal physics with complex few-body structures and dipolar gases.
The comparatively broad resonance located at about 180~G may be
a candidate for an entrance-channel dominated resonance \cite{Kohler2006poc,ChengRMP}.
Such a resonance could serve for the production of weakly bound molecules of halo character in a regime of large
scattering length \cite{ChengRMP,Ferlaino2008cbt} and to address
Efimov states composed of unequal mass particles \cite{Braaten2006uif}.

Dipolar gases of heteronuclear molecules in the rovibrational ground state have recently
attracted widespread attention because of the potentially strong intermolecular dipolar
force, which shows a long-range and anisotropic character \cite{Ni2008ahp,Goral2002qpd,Yi2007nqp}.
As recently demonstrated,
such a  system can be produced by first associating Feshbach molecules
and by then applying a coherent transfer scheme to populate the rovibrational
ground state \cite{Winkler2007cot,Danzl2008qgo,Lang2008utm,Ni2008ahp}.
Theoretical work \cite{Kotochigova2005air} has predicted strong dipolar forces
between RbCs molecules  in the rovibrational ground state.

\section*{Acknowledgments}
We thank S.\,Kokkelmans, S.\, Kotochigova, M.\,Goosen, and E.\,Tiesinga for stimulating
discussions and for sharing their insights in the Rb-Cs system.
We acknowledge support by the Austrian Science Fund (FWF) within SFB 15
(project part 16) and by the European Science Foundation (ESF)
in the framework of the EuroQUAM collective research project QuDipMol.
F.\,F.\,is supported within the Lise Meitner program of the
FWF.



%
%


\begin{references}
\bibitem{ED2}
The crossing angle is about $78^\circ$ for reason of optical access to the
trapping cell.
\bibitem{ED3}
We believe the shorter lifetime of Cs atoms to be due to the $6s\rightarrow 7s$
two-photon transition driven by the dipole trap light.
\bibitem{ED4}
In later experiments with optimized loading \cite{KarlThesis} we could
demonstrate BEC of Cs in this setup.
\bibitem{ED5}
For a typical cloud size of $\sigma=$20~$\mu$m, we estimate the broadening
caused by the field gradient $B^{'}$ to be $B^{'} \sigma\approx 70$~mG. The
temperature broadening can be estimated as $k_{\rm B} T/\Delta\mu$, where
$\Delta\mu$ is the relative magnetic moment between the atom pair state and the
molecular state causing the resonance. The broadening typically varies between
50 and 200~mG.
\bibitem{ED6}
We choose this two-photon transfer scheme because it is insensitive to magnetic
field fluctuations, whereas two subsequent one-photon transitions would be
strongly affected.
\bibitem{ED7}
A coherent transfer of Rb atoms  could not be successfully implemented our
setup. We believe the magnetic gradient field  in combination with the residual
magnetic-field dependence of the two-photon transition to be the main reason.
\end{references}

\end{document}